\documentclass[prb,twocolumn,showpacs]{revtex4}
\usepackage{graphicx}
\usepackage{psfrag}
\usepackage{amsmath}
\begin{document}

\title{The Random Field Ising Model on Hierarchical Lattices I:\\
Phase Diagram and Thermodynamics}
\author{Alexandre Rosas}
\author{S\'{e}rgio Coutinho}
\affiliation{Laborat\'orio de F\'\i sica Te\'orica e Computacional,\\
Universidade Federal de Pernambuco,\\
50670-901, Recife, Pernambuco, Brazil.}
  
\begin{abstract}
The phase diagram and the thermodynamics of the random field Ising model (RFIM) defined on a family of diamond hierarchical lattices of arbitrary dimension and scaling factor $b=2$ is investigated. The phase diagram is studied considering the flow of the renormalized joint probability distributions of couplings and fields. A continuous (Gaussian) and a discrete (delta-bimodal) initial symmetric probability distributions for the random fields with variance $H_{0}$ are particularly considered. The thermodynamics properties (energy, specific heat and magnetization) are obtained by an exact recurrence procedure and analyzed as function of the temperature and the random field. Close and above the paramagnetic-ferromagnetic transition evidences of the formation of rare but large field induced correlated clusters of reversed spins are observed. For all studied properties no strong qualitative distinct behavior is found whenever the continuous or the discrete distribution of random fields are considered. 
\end{abstract}
\pacs{61.43.-j, 64.60.Ak, 64.60.Fr}

\maketitle

\section{Introduction}
The study of random systems with quenched disorder deserved many experimental and theoretical efforts in the last quarter of century, but no significant progress has been achieved to full understanding its phase transition and critical behavior. In general the bond or the site disorder are considered to modeling real systems, the paradigms of these models being the Edwards-Anderson spin-glass model and random field Ising model (RFIM), respectively \cite{young97}. Regarding its critical behavior, the RFIM, which is believed to belong to the same class of universality as the diluted antiferromagnet under a uniform magnetic field \cite{fishman79}, is known to undergo a phase transition from a disordered phase to a long-range ferromagnetic phase when the temperature and strength of the random field are decreased below some critical values for systems with dimension greater than 2. However, the nature of such transitions remains not well understood even whether it is first-order or continuous. Despite many difficulties arising from non-equilibrium effects, recent experiments on magnetically diluted antiferromagnets, such us $Fe_{x}Zn_{1-x}F_{2}$, with high concentrations and under high fields, provide an experimental characterization of the equilibrium critical behavior of the 3d-RFIM \cite{belanger00}. The lack of exact solvable models beyond mean filed-like approximation has driven the recent theoretical efforts to use sophisticated numerical simulation approaches to study the temperature and field behavior of the energy, specific heat and magnetization of the RFIM \cite{machta00, sourlas99, hartmann01}. The phase diagram has been also subject to investigation by Monte Carlo simulation \cite{itakura00}. In this work we investigate the thermodynamic behavior of the RFIM defined on diamond hierarchical lattices (DHL's hereafter) with scaling factor $b=2$. 

Hierarchical lattices have been used, since long time ago, as a framework to study quenched disordered magnetic systems \cite{southern77}, as well as to study homogeneous \cite{tsallis98} and deterministic aperiodic \cite{andrade99} spin models. Whenever viewed as a scheme of real space renormalization approximation they offer only crude approximations for the critical properties of the corresponding homogeneous systems on Bravais lattices. However, for random systems like spin glasses, they give results comparable to other approaches for three-dimensional lattices, such as Monte Carlo simulations. It is worth to mention that the values of the critical temperature and of the critical exponents associated to the order parameter and correlation length for the Ising spin glass model on diamond hierarchical lattices with graph fractal dimension $d_F=3$ are surprisingly close to those obtained by numerical simulations for the model defined on the cubic lattice \cite{prakash97}. It seems that the randomness of couplings and/or fields washes out the effects caused by the lack of translational invariance and by the concomitant high inhomogeneous coordination-number distribution of the hierarchical lattices. On the other hand, when viewed as an exactly solvable model approach for random spin models, they can provide {\it exact} results to analyze the dependence of the order parameter on temperature and field. For instance, the local structures of the Edwards-Anderson order parameter of the short-range Ising spin-glass model defined on the DHL's has been analyzed by means of this methodology \cite{nogueira97}. Moreover, concerning the critical behavior of the latter model, it was possible to obtain strong evidences of absence of breaking of universality \cite{nogueira98,nogueira99} regarding the probability distributions of coupling constants, in contrast to what is claimed by some authors \cite{bernardi}. 

Here, we adapt to RFIM the methodology developed to the Ising spin-glass model under a field \cite{donato99}. Our aim in this paper is to study the phase diagram and the temperature and field dependence of the thermodynamic potentials (internal energy, specific heat and magnetization). The phase diagram is explored by the renormalization-group point of view. The thermodynamic properties are obtained from the site-to-site magnetization, which by it turn is calculated via a combined method encompassing real space renormalization-group approach and an exact recurrence procedure. We consider both a continuous (Gaussian) and a discrete (delta-bimodal) initial symmetric probability distributions for the random fields with variance $H_{0}$. In a subsequent paper, we focus our attention to the ground state critical properties by extending the present approach to the $T=0$ limit.

The remaining of this paper is organized as follow. In the next section we define the model and discuss the renormalization flow. In section~\ref{loc-mag} we analyze the local magnetization either with emphasis in the thermodynamics properties or in the local magnetization structure. Finally, in section~\ref{sec:conclusion}, we summarize our results.

\section{The model Hamiltonian and the Renormalization Flow}

\begin{figure}

\begin{center}
\resizebox*{8.5cm}{!}{\includegraphics{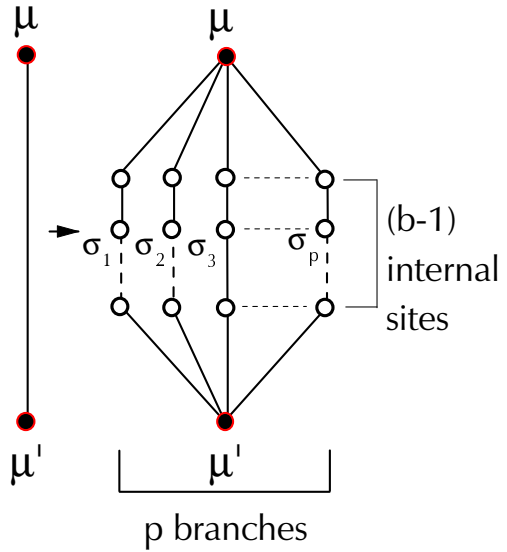}}
\caption{Schematic construction of a diamond hierarchical lattice with a basic unit composed by $p$ branches and $(b-1)$ internal sites (scaling factor b).}
\label{fig1}
\end{center}
\end{figure}

We consider the RFIM on a DHL with an arbitrary fractal dimension. This family of lattices is recursively constructed starting from a basic unit (called first generation) composed by two roots sites connected by a  set of $p$ parallel branches, each one formed by a series of $b$ connected bonds. Further generations are obtained by replacing all bonds of the previous one by the basic unit itself, as sketched in Figure~\ref{fig1}. The resulting lattice is a self-similar graph with the fractal dimension $d_{F}=1+\log p/\log b$. The number of sites and the number of bonds are respectively given by $N_{S}=2+(b-1)p[(bp)^{N}-1]/(bp-1)$ and $N_{B}=(bp)^{N}$, $N$ being the number of generations or hierarchies.

The RFIM energy is defined by:

\begin{equation}
{\cal H}=-\sum_{<ij>}J_{ij}\sigma_{i}\sigma_{j}-\sum _{i}H_{i}\sigma_{i},
\end{equation}
\noindent
where $J_{ij}=J_0>0$  are the ferromagnetic coupling constants of the pairs $<i,j>$ of  nearest neighbors spins, $\{H_{i}\}$ are the quenched local magnetic field at site $i$ and $\sigma_{i}=\pm 1$ are the corresponding Ising spin variables. The random field variables $\{H_i\}$ are chosen from a probability distribution $P(H)$ such that $[H^{2}]_{H}=H_{0}^{2}$ and $[H]_{H}=0$, $[...]_{H}$ meaning the average over the random field disorder. In the present work we consider the Gaussian and the delta-bimodal distributions given respectively by:

\begin{subequations}
\label{distribution}
\begin{align}
P(H_{i})&=\frac 1{\sqrt{2\pi}H_{0}}\exp (-\frac 12 \frac {H_{i}^2}{H_{0}^{2}}) \\   
P(H_{i})&=\frac 12\left[ \delta (H_{i}-H_{0})+\delta (H_{i}+H_{0})\right].
\end{align}
\end{subequations}

\begin{figure}

\begin{center}
\resizebox*{8.5cm}{!}{\includegraphics{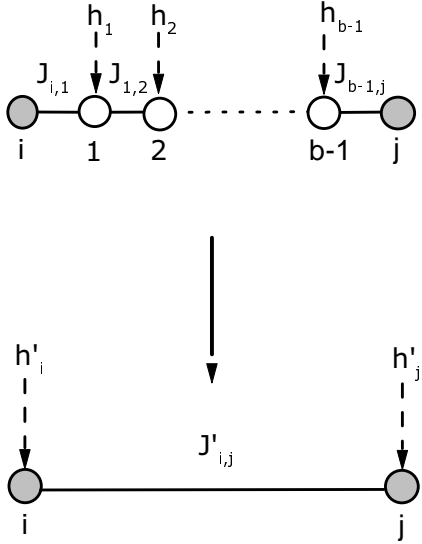}}
\caption{Renormalization process of a single branch to generate the joint probability distribution ${\cal P}(\Omega_{ij})$.}
\label{fig2}
\end{center}
\end{figure}

Building a N-generation lattice with uniform ferromagnetic configuration of couplings constants and a random configuration of local fields defined accordingly to a chosen distribution, one can decimate the spins variables introduced in the last generation obtaining a (N-1)-generation lattice with a new configurations of correlated effective couplings and local fields. These new coupling and fields are indeed locally correlated accordingly a joint probability distribution ${\cal P}(\Omega_{ij})$, where $\Omega_{ij}\equiv \{ J_{ij}, h_{i}, h_{j} \} $ represents the set of variables triplets associated with a given renormalized branch of a given basic unit of the previous generation after the decimation of its ($b-1$) internal sites, as sketched in the Figure \ref{fig2}. 

Nevertheless, to obtain the new random variables of the renormalized lattice one has to take into account the contributions coming from all decimated branches following the lattice topology. Each new coupling constant being just the sum of the renormalized coupling of each branch belonging to the corresponding previous basic unit, while the field variable associated with a given site being the sum of the renormalized fields coming from all branches (of all basic unities) arising from that site, in addition to the pre-existing field acting on that site. Such procedure weight correctly the sites accordingly to the high inhomogeneities found in the site coordination-number distribution along the lattice. The decimation process of a given basic unit gives rise to the following renormalization equation for the coupling and fields:

\begin{subequations}
  \label{fluxo}
  \begin{align}
    K' & = \frac{1}{4}\sum_{i=1}^{b} \left [ x_{1i} - y_{1i} \right . \nonumber \\
       & \left . + \log \frac{1+e^{r_{1i}-x_{1i}}+e^{-r_{1i}-x_{1i}}+e^{-2 x_{1i}}}{1+e^{s_{1i}-y_{1i}}+e^{-s_{1i}-y_{1i}}+e^{-2 y_{1i}}} \right ] 
  \end{align}
  \begin{align}
    h' & = h+\frac{1}{4}\sum_{i=1}^{b} \left [ x_{2i} - y_{2i} \right . \nonumber \\
       & \left . + \log \frac{1+e^{r_{2i}-x_{2i}}+e^{-r_{2i}-x_{2i}}+e^{-2 x_{2i}}}{1+e^{s_{2i}-y_{2i}}+e^{-s_{2i}-y_{2i}}+e^{-2 y_{2i}}} \right ] 
  \end{align}
  \begin{align}
     \bar{h}' & = \bar{h}+\frac{1}{4}\sum_{i=1}^{b} \left [ x_{3i} - y_{3i} \right . \nonumber \\
       & \left . + \log \frac{1+e^{r_{3i}-x_{3i}}+e^{-r_{3i}-x_{3i}}+e^{-2 x_{3i}}}{1+e^{s_{3i}-y_{3i}}+e^{-s_{3i}-y_{3i}}+e^{-2 y_{3i}}} \right ]
  \end{align}
\end{subequations}
where $K=\beta J$, $h_i=\beta H$ and the prime refering to the renormalized variables as usual. We have also defined
\begin{align*}
  x_{1i} & = \mathrm{max}(|2h_i|,2|K_{1i}+K_{2i}|) \\
  y_{1i} & = \mathrm{max}(|2h_i|,2|K_{1i}-K_{2i}|) \\
  r_{1i} & = \mathrm{min}(|2h_i|,2|K_{1i}+K_{2i}|) \\
  s_{1i} & = \mathrm{min}(|2h_i|,2|K_{1i}-K_{2i}|) \\
  x_{2i} & = \mathrm{max}(|2K_{1i}|,2|h_i+K_{2i}|) \\
  y_{2i} & = \mathrm{max}(|2K_{1i}|,2|h_i-K_{2i}|) \\
  r_{2i} & = \mathrm{min}(|2K_{1i}|,2|h_i+K_{2i}|) \\
  s_{2i} & = \mathrm{min}(|2K_{1i}|,2|h_i-K_{2i}|) \\
  x_{3i} & = \mathrm{max}(|2K_{2i}|,2|h_i+K_{1i}|) \\
  y_{3i} & = \mathrm{max}(|2K_{2i}|,2|h_i-K_{1i}|) \\
  r_{3i} & = \mathrm{min}(|2K_{2i}|,2|h_i+K_{1i}|) \\
  s_{3i} & = \mathrm{min}(|2K_{2i}|,2|h_i-K_{1i}|).
\end{align*}
Here, $\mathrm{max}(x,y)$ and $\mathrm{min}(x,y)$ are functions that return the maximum or minimum of their arguments respectively.

Equations \ref{fluxo} are rather general and were firstly deduced in the context of the short range Ising spin glass (SRISG) model under an external uniform magnetic field \cite{donato99}. Actually, for both systems, RFIM and SRISG model, equations \ref{fluxo} govern the evolution of the joint probability distribution of couplings and fields under the real space decimation scheme, the distinction between the two models being solely introduced by the choice of the initial distributions of coupling and fields, which determines the evolution towards the corresponding ``fixed-point'' distribution. Within the renormalization approach, the phase diagram of the RFIM can be directly determined by the overall behavior of the flow of the joint distribution under the renormalization process. To generate this flow we adopt the numerical reservoir procedure where pools of correlated coupling constants and fields $(\Omega_{ij})$ are build up from the equations \ref{fluxo}. For zero fields, this approach recovers the one often used to study the flow of the coupling constant distribution of the zero field SRISG model \cite{southern77}. In that case, it is worth to mention that the distribution evolves within the manifold defined by the trivial ``zero-field distribution'' (a delta-function at the origin), within the general space of the joint distribution the couplings and fields. For the RFIM case, we consider the initial distributions for the fields as given by equations \ref{distribution} and by a delta function $P(J_{i,j})=\delta(J_{i,j}-J)$, $J>0$, for the coupling constants. We analyze the evolution of the first  moment of the renormalized distributions of couplings as a function of the temperature and the variance $H_{0}$ of the initial distribution of random fields. The nature of the magnetic phase at a given temperature and field $H_{0}$ is determined by monitoring the evolution of $\bar J=[ J]_{J}$ and  and the field distribution variance $H=[H^2]^{1/2}$, for a $N$-generation lattice. As $N \rightarrow \infty$ the paramagnetic phase is stable provide $\lim_{N \rightarrow \infty} \bar J \rightarrow 0$ and $\lim_{N \rightarrow \infty} H \rightarrow \infty$, while the ferromagnetic phase is stable whenever $\lim_{N \rightarrow \infty} \bar J \rightarrow \infty$ and $\lim_{N \rightarrow \infty} \bar J/H \rightarrow \infty$.

\begin{figure}
  \psfrag{JH0}{$\bar J/H$}
  \psfrag{renormalizacao}{\hspace*{0.2cm} $N$}
\resizebox*{8.5cm}{!}{\includegraphics{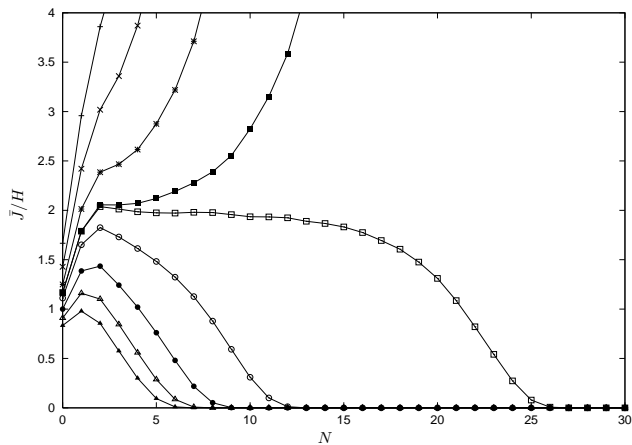}}
\caption{Typical example of a flow diagram for a $d_F=3$ DHL. For a Gaussian distribution with $T=0.8J/k_B$, the transition field can be determined as the one for which the behavior of $\bar J/H$ changes from the paramagnetic to the ferromagnetic behavior. The following fields are plotted from the top to the bottom of the graphic $0.6$, $0.7$, $0.8$, $0.856$, $0.858$, $0.9$, $1.0$, $1.1$ and $1.2$.}
\label{fig3}
\end{figure}

\begin{figure}
  \psfrag{H0J0}{\Large $H_0/J_0$}
  \psfrag{TJ0}{\Large $k_B T/J_0$}
  \resizebox*{8.5cm}{!}{\includegraphics{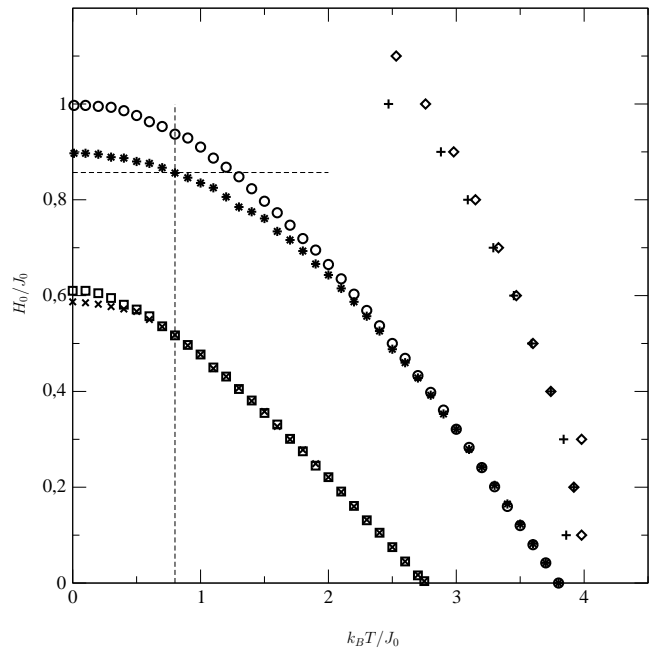}}
  \caption{Phase diagram $H_{0}/ \bar J_{0}$ against $T/ \bar J_{0}$, for a $d_F=3$, open circle (Gaussian) and stars (delta-bimodal), and $d_F=2.58\ldots$, open squares (Gaussian) and $\times$ (delta-bimodal), DHL. The open diamonds (Gaussian) and the crosses (delta-bimodal) are an approximate boundary for the region where the random field act strongly.}
  \label{phase}
\end{figure}

To obtain the phase diagram of the RFIM on a $d_F=3$ lattice we consider a diamond hierarchical lattice with $b=2$ and $p=4$ and analyze the evolution of the joint distribution accordingly to Eq. \ref{fluxo}. In figure \ref{fig3} we show a typical plot of the evolution of $\bar J/H$ for a fixed temperature and several values of $H_0$ up to $N=30$ renormalization steps. The phase boundary can be traced in the diagram $H_{0}/ J_{0}$ against $k_BT/ J_{0}$, as shown in figure \ref{phase} for the $d_{F}=3$ and the $d_F=2.58\ldots$ hierarchical lattices. There are slight differences at low temperatures when the plots obtained by considering the Gaussian or the delta-bimodal distribution of fields are compared, giving rise to distinct zero critical fields for the Gaussian ($H_0^c\simeq 1 J$) and delta-bimodal  ($H_0^c=\simeq 0.9 J$) models. However, both plots give the exact zero field critical temperature for the pure ferromagnetic Ising model on this lattice given by $k_{B}T_{C}/J=1.38...$ for the three dimensional lattice. In Fig~\ref{phase} we also plot boundary for the region where the action of the random field is more remarkable. This boundary is determined and discussed below in section~\ref{sec:localmag}.

It is worth to mention that we have also performed a phase diagram for the $d_F=2$ DHL, but no ferromagnetic phase was observed -- even for very small fields, the renormalization flow goes to the paramagnetic phase.

\section{The local magnetization}\label{loc-mag}
The method for evaluation of the local magnetization is based on the equivalence of the partition function of the model Hamiltonian for a lattice with $N$ generations with the one of a single basic unit introduced in the $N^{th}$ step with effective fields acting on the spins of its root sites in addition to an effective interaction coupling these spins. These unknown local effective interaction and fields mimic the influence of the remaining lattice spins couplings transmitted by the root spins of that unit. Actually, these effective interaction and fields arise when we perform the trace over the spins variables of the remaining lattice considering as a decoration unit coupling the root sites in the spirit of the decoration transformation \cite{fisher51}. The local magnetization and the pair correlation function of all sites of the basic unit of the effective system can be calculated as a function of its coupling constants and local fields and the unknown effective coupling and fields. From a subset of these equations, the unknown variables can be calculated as a function of the local magnetization and the pair correlation function involving the internal sites. Now, the unknown effective fields and couplings can be eliminated ending up to coupled recursive relations giving the local magnetization and the pair correlation functions involving the internal sites in terms of the corresponding values involving the root sites. This approach has been extensively applied to investigate the local order parameter of several magnetic models defined on hierarchical lattices. Those studies comprise the ferromagnetic Ising model with uniform \cite{morgado90,morgado91} and aperiodic interactions \cite{nogueira2000}, the spin-glass Ising model \cite{nogueira97, nogueira98} and the ferromagnetic Potts\cite{ladario} model defined on diamond hierarchical lattices,  as well as the Ising model (both the pure and the spin-glass cases) defined on the Wheatstone bridge hierarchical lattice \cite{camelo99} and on the m-sheet Sierpinskii Gasket fractal lattice \cite{lima99}. 

For the RFIM the coupled recursive equations between local magnetization and pair correlation functions are written as 

\begin{subequations}
  \label{recurrence}
  \begin{align}
    <\sigma> & = A_1 + A_2 <\mu_1> \nonumber \\
             & + A_3 <\mu_2> + A_4 <\mu_1\mu_2> \label{magnetization}\\
    <\sigma \mu_1> &= A_2 + A_1 <\mu_1> \nonumber \\
             & + A_4 <\mu_2> + A_3  <\mu_1\mu_2> \\
    <\sigma \mu_2> &= A_3 + A_4 <\mu_1> \nonumber \\
             & + A_1 <\mu_2> + A_2 <\mu_1\mu_2>
  \end{align}
\end{subequations}
where $\sigma$ is the magnetization of the internal site of a given branch, $\mu_1$ and $\mu_2$ are the magnetizations of the root sites of this branch, $h$, $K_1$ and $K_2$ and the interaction constants acting on $\sigma$. The coefficients of equations~(\ref{recurrence}) are given by 

\begin{subequations}
  \begin{align}
    4A_1&=\tanh (h+K_1+K_2) + \tanh (h+K_1-K_2) \notag \\
    &+ \tanh (h-K_1+K_2) +\tanh (h-K_1-K_2), \\
    4A_2&=\tanh (h+K_1+K_2) + \tanh (h+K_1-K_2) \notag\\
    &- \tanh (h-K_1+K_2) - \tanh (h-K_1-K_2), \\
    4A_3&=\tanh (h+K_1+K_2) - \tanh (h+K_1-K_2) \notag \\
    &+ \tanh (h-K_1+K_2) - \tanh (h-K_1-K_2), \\
    4A_4&=\tanh (h+K_1+K_2) - \tanh (h+K_1-K_2) \notag \\
    &- \tanh (h-K_1+K_2) + \tanh (h-K_1-K_2).
  \end{align}
\end{subequations}

For zero field systems, $A_3=A_4=0$ and equations \ref{recurrence} decouple. In this case, the equation \ref{magnetization} was used to study the critical and multifractal properties of the Edwards -Anderson order parameter of the SRISG model \cite{nogueira97,nogueira99} and the magnetization of the pure ferromagnetic Ising model \cite{morgado90,morgado91}

To obtain the site-to-site magnetization and the bond-to-bond correlation functions of the whole lattice we proceed the following script: firstly, we consider a $N$ generation lattice and successively decimate the spins up to the first generation (basic unit), keeping track the corresponding local values of couplings and fields variables $(\Omega_{ij})$ at each step. Then, we calculate the initial values for the magnetization and pair correlation function of the root sites, using the effective partition function. Now, we recursively calculate the local magnetization $<\sigma_i>$ and the bond correlation function $<\sigma_i \mu>$ using at each step the exactly stored local values for $\Omega_{ij}$. This procedure requires large computational efforts as the number of sites and bonds grow exponentially of with the number of generations $N_{S} \sim N_{B} \sim b^{N}$. Moreover, one has to average over many realizations of the initial distributions $P(H)$ (samples), yielding to the magnetization per spin and the dimensionless internal energy per spin straightforwardly obtained by 
\begin{equation}
\label{magtotal}
m=\frac{1}{N_{S}}\sum_{i} [< \sigma_i >]_{c}
\end{equation}
\begin{equation}
\label{energy}
E=\frac{1}{N_{S}} \{ \sum_{<i,j>} [< \sigma_i \sigma_j >]_{c}+
\sum_i [h_i <\sigma_i>]_{c} \}
\end{equation}
\noindent
where $[...]_{c}$ denotes the configurational average taken over many samples and 
$h_i=H_i/J_0$.

\subsection{Thermodynamic properties}
\begin{figure}
    \psfrag{M}{\Large $M$}
    \psfrag{H}{\Large $H_0/J$}
    \psfrag{T}{\Large $k_BT/J$}
    \resizebox*{8.5cm}{!}{\includegraphics{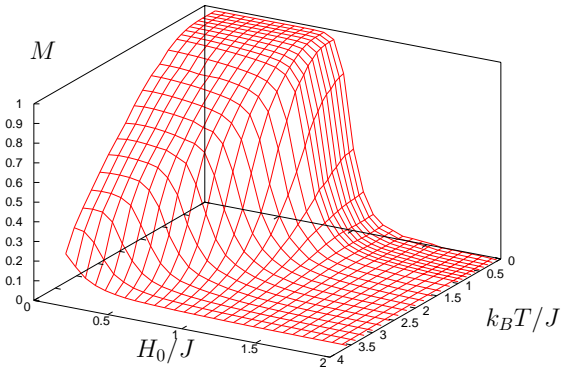}}
    \caption{Average magnetization as a function of the temperature and the strength of the random field for the RFIM with a delta bimodal distribution and $d_F=3$.}
\label{fig5}
\end{figure}

\begin{figure}
    \psfrag{M}{\Large $M$}
    \psfrag{H}{\Large $H_0/J$}
    \psfrag{T}{\Large $k_BT/J$}
    \resizebox*{8.5cm}{!}{\includegraphics{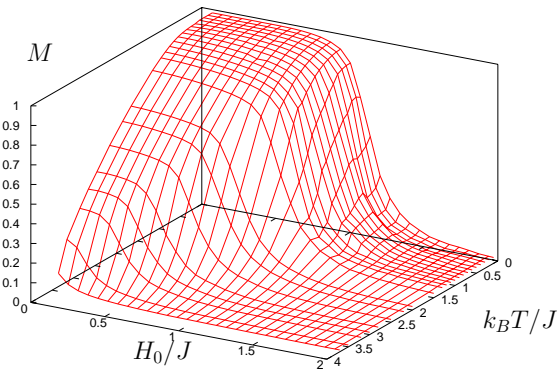}}
    \caption{ Average magnetization as a function of the temperature and the strength of the random field for the RFIM with a Gaussian distribution $d_F=3$.}
\label{fig6}
\end{figure}

\begin{figure}
    \psfrag{E}{$E$}
    \psfrag{H}{$H_0/J$}
    \psfrag{T}{$k_BT/J$}
    \resizebox*{8.5cm}{!}{\includegraphics{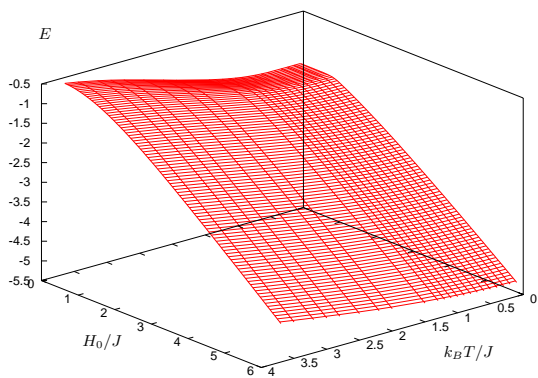}}
    \caption{Internal energy as a function of the temperature and the strength of the random field for the RFIM with a Gaussian distribution and $d_F=3$.}
\label{fig7}
\end{figure}

\begin{figure}
    \psfrag{C}{$C$}
    \psfrag{H}{$H_0/J$}
    \psfrag{T}{$k_BT/J$}
    \resizebox*{8.5cm}{!}{\includegraphics{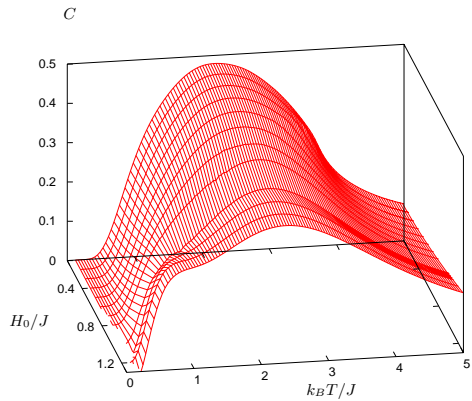}}
    \caption{Specific heat as a function of the temperature and the strength of the random field for the RFIM with a delta-bimodal distribution and $d_F=3$.}
    \label{fig8}
\end{figure}

\begin{figure}
    \psfrag{C}{$C$}
    \psfrag{H}{$H_0/J$}
    \psfrag{T}{$k_BT/J$}
    \resizebox*{8.5cm}{!}{\includegraphics{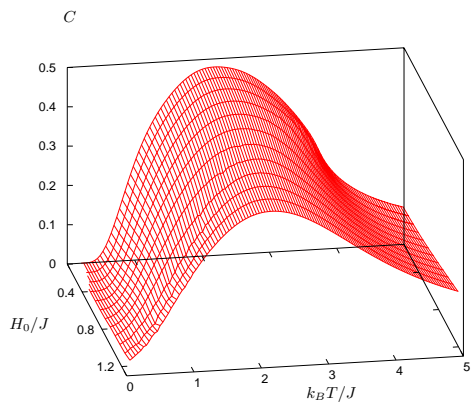}}
    \caption{Specific heat as a function of the temperature and the strength of the random field for the RFIM with a Gaussian distribution and $d_F=3$.}
    \label{fig9}
\end{figure}

In figures \ref{fig5} to \ref{fig9}, we present the temperature and field
dependence of the magnetization, the internal energy and the specific heat, obtained from equations \ref{magtotal} and \ref{energy} -- the specific heat being calculated by the numerical derivative of the internal energy with respect to the temperature. In each figure, we show the results for either the Gaussian or the delta-bimodal probability distributions calculated by the exact methodology considering a $d_F=3$ DHL with $N=7$ hierarchies which correspond to $N_S \sim 1.2 e+6$ and $N_B \sim 2.1 e+6$ sites and bonds respectively.

In Fig.~\ref{fig5} and \ref{fig6}, whatever is the probability distribution, the magnetization decreases faster for high fields (below the critical fields) as the temperature is increased, indicating that stronger is the field greater is the probability of appearance of correlated clusters of reversed spins with respect to the magnetization. A detailed analysis of the distribution of reversed clusters, at zero temperature, as a function of the random field and an external uniform field is now under consideration and the results will be present elsewhere. Figures \ref{fig5} and \ref{fig6} also show that the field and temperature dependence are qualitatively similar with respect to the random field initial distribution. In the subsequent paper, we analyze the dramatic behavior of the magnetization at $T=0$, suggesting a discontinuity with $\beta=0$ exponent.

In figure Fig.~\ref{fig7}, the behavior of the internal energy as function of the temperature and field strength is presented for the Gaussian model (the delta-bimodal one being very similar). The internal energy increases with the increasing of the temperature for a fixed value of the field the slope becoming bigger (smaller) for values of the field below (above) $H_{0}^{c}(T)$. For small temperatures and fields, the internal energy is dominated by the exchange term, so that it is almost constant in this region. For high fields, however, it is the local field term that dominates it, thus justifying the linear dependence the internal energy upon $H_0$, for $H_0\gg H_0^c$.

The specific heat behavior is displayed in Fig.~\ref{fig8} for the delta-bimodal model and in Fig.~\ref{fig9} for the Gaussian one. We notice that the temperature where the maximum occurs almost does not depend upon $H_0$, and consequently it does not coincide with the corresponding transition temperature $T_c(H_0)$. This behavior is characteristic of Ising spin systems defined on the diamond family of hierarchical lattices with $d_F >2$ \cite{donato99, donato96}. For fields near and above the critical one, $H_0(0)$, we observe that the specific heat for the delta-bimodal model displays a ``plateau'', absent in the Gaussian model. This difference appears only in the specific heat since, among the studied thermodynamic potential, it is the most susceptible to fluctuations.

\subsection{Local magnetization structure}\label{sec:localmag}
\begin{figure}
    \psfrag{qea}{$q_{EA}$}
    \psfrag{H}{$H_0/J$}
    \psfrag{T}{$k_BT/J$}
    \resizebox*{8.5cm}{!}{\includegraphics{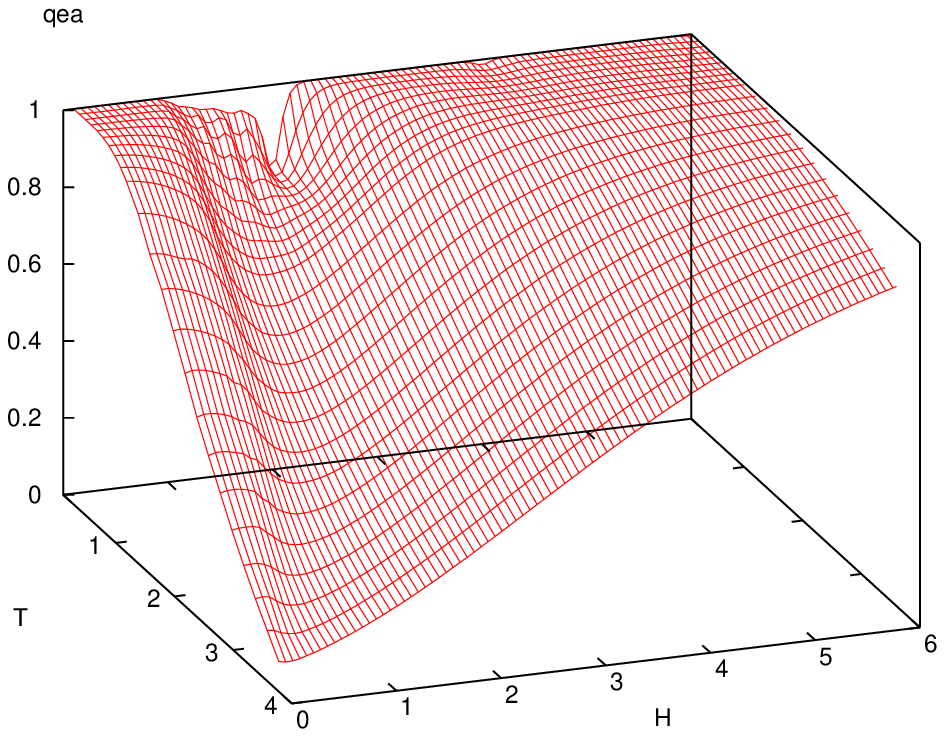}}
    \caption{Edwards-Anderson parameter as a function of the temperature and the strength of the random field for the RFIM with a delta-bimodal distribution and $d_F=3$.}
    \label{fig10}
\end{figure}

\begin{figure}
    \psfrag{qea}{$q_{EA}$}
    \psfrag{H}{$H_0/J$}
    \psfrag{T}{$k_BT/J$}
    \resizebox*{8.5cm}{!}{\includegraphics{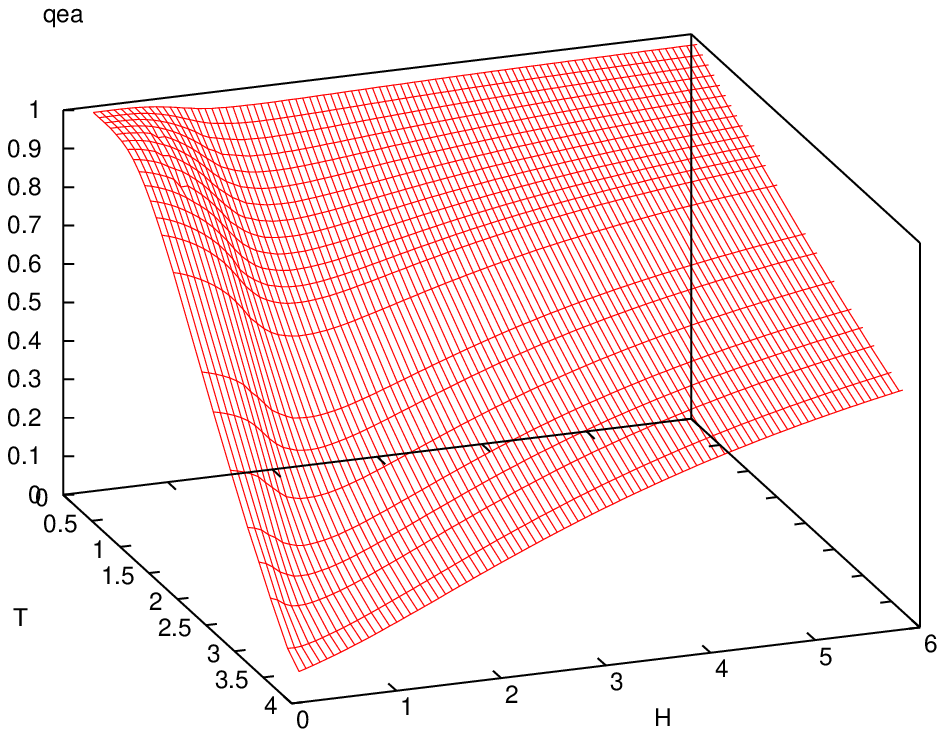}}
    \caption{ Edwards-Anderson parameter as a function of the temperature and the strength of the random field for the RFIM with a delta-bimodal distribution and $d_F=3$.}
    \label{fig11}
\end{figure}

In this section, we focus our attention on the structure of the local magnetization, in order to analyze the effect of spin reversion. The Edwards-Anderson parameter $q_{EA} = 1/N_S \sum_i [<\sigma_i>^2]_c$ is calculated and shown in figures \ref{fig10} and \ref{fig11}, for the delta-bimodal and the Gaussian model respectively. In both cases, there is a depression the transition for low temperatures and field near and above the critical one. This depression is a consequence of the competition between the exchange and the field terms, which in the transitions are almost equal, giving rise to small local magnetization. It is worth to notice that the depression is more evident in the delta-bimodal model, specially for low temperatures, since the probability of exact frustrated spins are non zero in this model. For higher fields, the spins follow their local fields and consequently $q_{EA} \rightarrow 1$. Analogously, for small fields and temperatures (ferromagnetic phase), all spins are correlated and aligned, so that we also have  $q_{EA} \rightarrow 1$. 

\begin{figure}
    \psfrag{D}{$\Delta$}
    \psfrag{H}{$H_0/J$}
    \psfrag{T}{$k_BT/J$}
    \resizebox*{8.5cm}{!}{\includegraphics{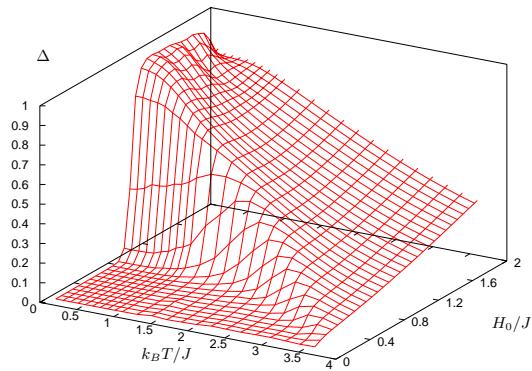}}
    \caption{Variance of magnetization distribution as a function of the temperature and the strength of the random field for the RFIM with a delta-bimodal distribution and $d_F=3$ .}
\label{fig12}
\end{figure}

\begin{figure}
    \psfrag{D}{$\Delta$}
    \psfrag{H}{$H_0/J$}
    \psfrag{T}{$k_BT/J$}
    \resizebox*{8.5cm}{!}{\includegraphics{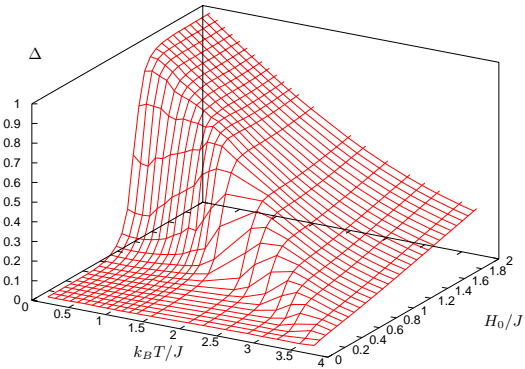}}
    \caption{Variance of magnetization distribution as a function of the temperature and the strength of the random field for the RFIM with a Gaussian distribution and $d_F=3$.}
    \label{fig13}
\end{figure}

In order to investigate the role of the reversed clusters we analyze the temperature and field dependence of the configurational average of the magnetization distribution $P_{l}(m)$ (the $l$-index labeling the samples) variance, that is,
\begin{equation}
\Delta =[\int dm m^2 P_{l}(m)]-[\int dm m P_{l}(m)]^2.
\end{equation} \\
\noindent
$\Delta$ gives information about the average numbers of site with reversed magnetization induced by the random field. In figures \ref{fig12} and  \ref{fig13} we plot $\Delta$ as function of the temperature and field, for the Gaussian and the delta-bimodal models, respectively. For a fixed temperature, the number of sites with reversed magnetization increases with $H_0$, the beginning of the increasing region being determined by the corresponding critical field $H_0^c(T)$. Looking at figures \ref{fig12} and  \ref{fig13} for fixed fields instead, we notice that $\Delta$ displays a maximum above which the temperature dominates the system behavior and all the local magnetizations vanishes. We also observe that the maxima shifts for low temperatures as $H_0$ increases. This is a consequence of the decreasing of the critical temperatures with the field strength. Since the maximum of $\Delta$ determines the point where the effect of the random field is the strongest, we determine the boundary of the region where the field ``plays its role'' as the point that $\Delta$ is half of its maximum. This boundary should be related with the experimentally observed non-equilibrium line\cite{montenegro99}. We also point out that for very low temperatures and values of $H_0$ close and above $H_{0}^c$ one observes a small increase of the $\Delta$ parameter, indicating the persistence of sites with reversed magnetization at ground sate (this behavior is more evident for the Gaussian model). 

\section{Conclusion}
\label{sec:conclusion}

We studied  the RFIM on DHL with fractal dimension three, considering either a Gaussian or a delta-bimodal probability distribution for the random fields. We analyzed the renormalization flow as well as thermodynamic properties. We obtained the phase diagram for the three dimensional and for the $d_F=2.58\ldots$ systems. We also obtained a boundary line separating the region dominated by the random fields from the one dominated by the temperature, for the three dimensional system. Concerning to the thermodynamics, we observed that the thermodynamic potentials obtained within both probability distributions displays a very similar behavior, the Edwards-Anderson parameter at low temperatures (as a consequence of exact frustrated sites) and the heat capacity near the critical field (by virtue of strong fluctuations) being the exceptions. The parameter $\Delta$, which gives information about the clusters of reversed spins, gives useful information of the random field effect, displaying useful information about the competition among the random field, temperature and the ferromagnetic order.

\acknowledgments

We thank to F. C. Montenegro for helpful discussions. AR is greatful to CNPq and FACEPE (Brazilian granting agencies) for financial support. This work also received financial support from FINEP (under the grant PRONEX 76.97.1004.00), CNPq and CAPES.

\end{document}